\documentclass[aps,prb,floatfix,twocolumn,
      superscriptaddress,showpacs,amsmath,amssymb]{revtex4}
\usepackage{graphicx}
\usepackage{epsfig,color} 
\allowdisplaybreaks

\newcommand{\be}{\begin{equation}}
\newcommand{\ee}{\end{equation}}
\newcommand{\bea}{\begin{eqnarray}}
\newcommand{\eea}{\end{eqnarray}}
\newcommand{\ba}{\begin{eqnarray*}}
\newcommand{\ea}{\end{eqnarray*}}
\newcommand{\dagga}{{\phantom{\dagger}}}

\newcommand{\bk}{\mathbf{k}}

\newcommand{\dis}{\displaystyle}

\newcommand{\fract}[2]{\frac{\dis #1}{\dis #2}}
\newcommand{\Tr}{\mathrm{Tr}}
\newcommand{\eqn}[1]{(\ref{#1})}
\newcommand{\ket}[1]{\mid\! #1\rangle}
\newcommand{\bra}[1]{\langle #1\!\mid}

\newcommand{\bw}{\begin{widetext}}
\newcommand{\ew}{\end{widetext}}

\newcommand{\expv}[1]{\langle #1 \rangle}

\begin{document}

\title{$Z_2$-gauge theory description of the Mott transition in infinite dimensions} 

\author{Rok {\v Z}itko}

\affiliation{Jo{\v z}ef Stefan Institute, Jamova 39, SI-1000 Ljubljana, Slovenia} 
\affiliation{Faculty of Mathematics and Physics, University of Ljubljana, Jadranska 19, SI-1000 Ljubljana, Slovenia}

\author{Michele Fabrizio} 
\affiliation{International School for
  Advanced Studies (SISSA), and CNR-IOM Democritos, Via Bonomea
  265, I-34136 Trieste, Italy}

\date{\today} 


\begin{abstract}
The infinite dimensional half-filled Hubbard model can be mapped
exactly with no additional constraint onto a model of free fermions
coupled in a $Z_2$ gauge-invariant manner to auxiliary Ising spins in
a transverse field. In this slave-spin representation, the zero-temperature
insulator-to-metal transition translates into spontaneous breaking of
the local $Z_2$ gauge symmetry, which is not forbidden in infinite
dimensions, thus endowing the Mott transition of an order parameter
that is otherwise elusive in the original fermion representation. We
demonstrate this interesting scenario by exactly solving the effective
spin-fermion model by dynamical mean-field theory both at zero and at
finite temperature. 
\end{abstract}

\maketitle

\section{Introduction}

Sixty years after Sir Nevill Mott first envisaged it,~\cite{Mott} the
interaction driven metal-to-insulator (Mott's) transition remains an
object of considerable efforts aiming at its full comprehension. The
main difficulty resides in the inability of approximate methods based
on independent particles to properly capture a transition
that involves only part of electrons' degrees of freedom: their
charge. This is frustrating since the common attitude is tracing back
complex many-body phenomena to independent particle models, an
approach on which many successful techniques are based, e.g., the
mean-field theory or the density functional theory within the local
density approximation.

A tool that has been often employed to disentangle charge from
other electron quantum numbers, thus making Mott's localization
accessible already at the mean-field level, is to associate local
charge configurations to novel fictitious degrees of freedom. This
requires supplemental local constraints in order to project the
enlarged Hilbert space onto the physical one. 

A well-known implementation of this idea is the {\sl slave boson}
technique.~\cite{Barnes-slave-boson,Coleman-slave-boson} The
slave-boson mean-field theory is indeed able to describe the
metal-insulator transition.~\cite{Kotliar&Ruckenstein} In this
language, however, the Mott transition is associated with a local
$U(1)$ gauge symmetry that is maintained in the insulator but
spontaneously broken in the metal, which is physically not possible.
Restoring the symmetry to assess how robust are the mean-field results requires adding quantum fluctuations, which is not an easy
task.~\cite{Read-quantum-fluctuations,Bickers-RMP,Arrigoni}

In a recently proposed alternative route the slave bosons are
replaced by slave Ising
variables.~\cite{De-Medici,Z2-1,Z2-2,Marco-PRB} The advantage is that
the size of the enlarged local Hilbert space is only a few times
bigger than the physical one, not infinitely as in the slave-boson
approach. In addition, the continuous $U(1)$ local gauge symmetry 
is replaced by a discrete $Z_2$ one.
 
Let us consider the single-band model
\bea
\mathcal{H} \!&=&\! - \sum_{i,j}\sum_\sigma\,
t_{ij}\,c^\dagger_{i\sigma}c^\dagga_{j\sigma}
+ \sum_i\,\fract{U_i}{4}\,\bigg[2\big(\hat{n}_i-1\big)^2-1\bigg],\;\;\;\label{Ham}
\eea
where $\hat{n}_i=\sum_\sigma c^\dagger_{i\sigma}c^\dagga_{i\sigma}$, and, for later convenience,   
we added a constant term. 
The Hamiltonian $\mathcal{H}$ corresponds to the conventional Hubbard
model when $U_i=U$, $\forall\,i$, and to the Anderson (Wolff) impurity
model when $U_i=U$ for $i=i_0$ while $U_i=0$ for any $i\not = i_0$, with
the impurity sitting at $i_0$. The partition function $Z$
corresponding to the Hamiltonian in Eq.~\eqn{Ham} can be
shown~\cite{Marco-PRB} to be equivalent to 
\bea Z &=&
\Tr\Big(\text{e}^{-\beta\mathcal{H}}\Big) \equiv Z_*=
\Tr\Big(\text{e}^{-\beta\mathcal{H}_*}\;\prod_i\mathbb{P}_i\Big),\;\;\;\;\label{mapping}
\eea
where the slave-spin Hamiltonian $\mathcal{H}_*$ is
\bea \mathcal{H}_* &=& -
\sum_{i,j}\sum_\sigma\, t_{ij}\;\sigma^x_i\sigma^x_j\;
c^\dagger_{i\sigma}c^\dagga_{j\sigma} 
-\sum_i\,\fract{U_i}{4}\,\sigma^z_i,\label{H_*} \eea
with the Pauli matrices $\sigma^a_i$, $a=x,y,z$, describing auxiliary Ising spins, and the
projection operator \be \mathbb{P}_i = \frac{1}{2} +
\fract{\sigma^z_i\,\Omega_i}{2},\label{P_i} \ee with $\Omega_i$
defined as [see Eq.~\eqn{Ham}]
\be 
\Omega_i = 1 - 2\big(\hat{n}_i-1\big)^2.\label{Omega_i}
\ee 
We observe that $\Omega_i$ has eigenvalue $+1$ when $\hat{n}_i=1$ and
$-1$ when $\hat{n}_i=0$ or $2$; this operator thus corresponds to
charge fluctuations away from single occupancy of site $i$.
Therefore $\mathbb{P}_i$ is actually a projector onto the physical
Hilbert space defined by $\sigma^z_i=\Omega_i$. As anticipated, the
Hamiltonian $\mathcal{H}_*$ possesses a local $Z_2$ gauge symmetry \be
\sigma^x_i \to s_i \,\sigma^x_i,\qquad c^\dagger_{i\sigma}
\to
s_i\,c^\dagger_{i\sigma},
\label{Z2}
\ee
where $s_i=\pm 1$.

The equivalence of partition functions in Eq.~\eqn{mapping} was proved
by equating the perturbative expansion in $U_i$ of $Z$ with that of 
$Z_*$.~\cite{Marco-PRB} It was observed that the two terms that define
the projector $\mathbb{P}_i$ in Eq.~\eqn{P_i} have a very distinct
role in perturbation theory.~\cite{Marco-PRB} Specifically, the
first term $1/2$ accounts for all diagrams in the perturbation series
where $U_i$ is applied an even number of times, while the second term
$\sigma^z_i\,\Omega_i/2$ takes care of all diagrams where $U_i$ is
instead applied an odd number of times. 

This simple observation has a remarkable outcome. If $Z$ is known to
be an even function of some $U_i$, then the term
$\sigma^z_i\,\Omega_i/2$ in Eq.~\eqn{P_i} plays no role and can be
discarded. This result can be easily proven even without resorting to
the perturbation theory. If $Z(U_i)=Z(-U_i)$, also
$Z_*(U_i)=Z_*(-U_i)$. Since
$\sigma^x_i\,\mathcal{H}_*(U_i)\,\sigma^x_i = \mathcal{H}_*(-U_i)$ 
and $\sigma^x_i\,\mathbb{P}_i\,\sigma^x_i = 1-\mathbb{P}_i$, because
the partition function is invariant under a unitary transformation  
it follows that 
\bea Z_*\left(U_i,\mathbb{P}_i\right) &=&
Z_*\left(-U_i,\mathbb{P}_i\right) =
Z_*\left(U_i,1-\mathbb{P}_i\right)\\ 
&=&
Z_*\left(U_i,\fract{\mathbb{P}_i +
\left(1-\mathbb{P}_i\right)}{2}\right) = 
Z_*\left(U_i,\frac{1}{2}\right).\nonumber \eea

The fortunate event $Z(U_i)=Z(-U_i)$ occurs, in particular, at 
particle-hole symmetry in the Anderson impurity model.  Moreover,
through the well known dynamical mean-field theory (DMFT) mapping
between an Anderson impurity model and a Hubbard model on a lattice
with infinite coordination number,~\cite{DMFT} it follows that also
the latter possesses this property at particle-hole symmetry. 
In other words, the partitions functions of both models at the
particle-hole symmetric point can be calculated through the partition
function of the corresponding spin-fermion Hamiltonian
$\mathcal{H}_*$ without any constraint. 

This mapping was exploited for
the Anderson impurity model in Ref.~\onlinecite{Pierpaolo}, where the
equivalent spin-fermion Hamiltonian was exactly solved by the
numerical renormalization group (NRG)~\cite{Wilson_RMP, Pruschke_RMP},
unveiling the remarkable fact that the mean-field breaking of the
local $Z_2$ gauge symmetry is not spurious: it actually occurs in the
exact solution. In the same work it was also argued that a similar
phenomenon might occur even in the Hubbard model on
infinite-coordination lattices, where Elitzur's theorem~\cite{Elitzur}
does not apply hence a local $Z_2$ gauge symmetry can be spontaneously
broken.~\cite{Maslanka} 

Our aim here is just to verify this conjecture and examine all its
consequences. As a byproduct, the numerically exact solution of the
lattice model Eq.~\eqn{H_*} provides for the first time the chance to
assay how reliable is the mean-field approximation when applied to
slave-spin models, at least in the case of lattices with
infinite coordination number.

\begin{figure}[t]
\centering
\includegraphics[clip,width=0.48\textwidth]{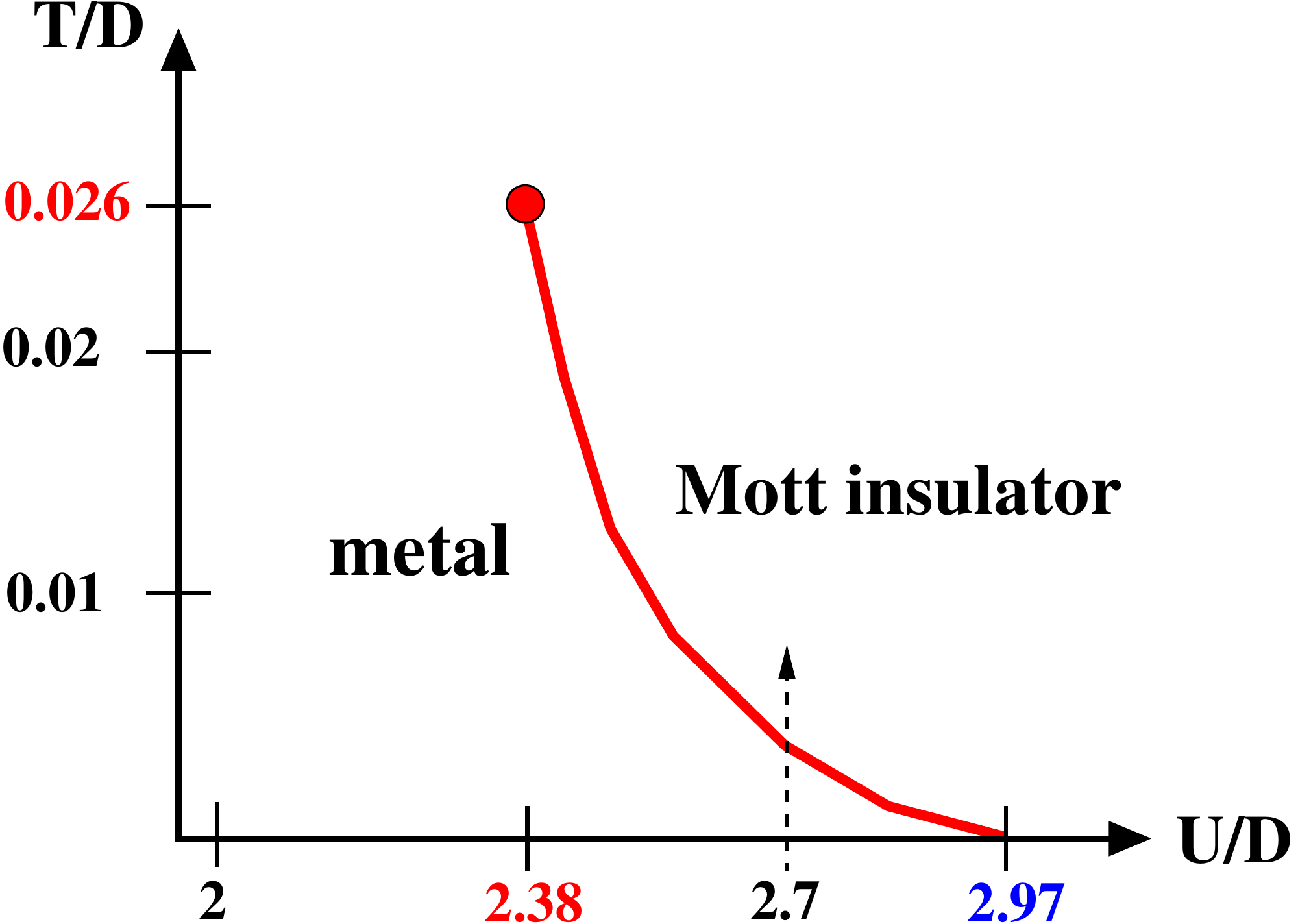}
\caption{(Color online) Sketch of the phase diagram of the
paramagnetic half-filled Hubbard model on a Bethe
lattice.~\cite{DMFT-finite-T} $D=2t$ is half the bandwidth. At
$U/D=2.7$, the first-order metal-insulator transition occurs at the
temperature $T \approx 0.0037D$. }
\label{phasediagram}
\end{figure}

\section{The model and its mean-field solution}
\label{Sec:mean-field}

The model we shall consider is the simple half-filled Hubbard model 
\bea
\mathcal{H}_\text{Hub} &=& -\frac{t}{\sqrt{z}} \sum_{<i,j>}\sum_\sigma\,
\big(c^\dagger_{i\sigma}c^\dagga_{j\sigma}+H.c.\big)\nonumber\\
&&
+ \frac{U}{4}\, \sum_i\,\Big[2\big(n_i-1\big)^2-1\Big],\label{Ham-Hub}
\eea
with nearest-neighbor hopping on a Bethe lattice in the limit of
infinite coordination number, $z\to\infty$. We focus on the fully
frustrated case where the antiferromagnetic order does not appear and
there is a Mott transition from a paramagnetic metal to a
paramagnetic insulator. The phase diagram of the model is
sketched in Fig.~\ref{phasediagram}.~\cite{DMFT-finite-T} The
transition line between the metal and the Mott insulator is first
order at any $T>0$ with a critical endpoint at $T_c \simeq
0.026\,D$,~\cite{Marcelo-DMFT} where $D=2t$ is half the bandwidth.

As mentioned, the partition function of the Hubbard model
$Z_\text{Hub}$ can be also calculated through 
\be
Z_\text{Hub} = \left(\fract{1}{2}\right)^N\;
\Tr\Big(\text{e}^{-\beta\mathcal{H}_\text{s-f}}\Big),\label{mapping-2}
\ee
where $N$ is the number of sites and $\mathcal{H}_\text{s-f}$ is the
spin-fermion Hamiltonian, Eq.~\eqn{H_*}, with $U_i=U$, $\forall\, i$,
namely 
\bea
\mathcal{H}_\text{s-f} &=& - \frac{t}{\sqrt{z}}\,\sum_{<i,j>}\sum_\sigma\,
\sigma^x_i\sigma^x_j\; \big(c^\dagger_{i\sigma}c^\dagga_{j\sigma}+H.c.\big)
\nonumber\\
&& -\fract{U}{4}\,\sum_i\,\sigma^z_i.\label{H_s-f}
\eea
It follows that the spin-fermion model Eq.~\eqn{H_s-f} must also have
the same phase diagram shown in Fig.~\ref{phasediagram}. 

The Hamiltonian $\mathcal{H}_\text{s-f}$ is invariant under the $N$ mutually
commuting unitary transformations $\sigma^z_i\,\Omega_i$, $\forall\,
i$. Therefore we can divide the full Hilbert space into $2^N$ orthogonal subspaces identified by the sets of bits $\{n\} =
(n_1,n_2,\dots,n_i,\dots,n_N)$, with $n_i=0,1$, which contain states
such that
\be
\sigma^z_i\,\Omega_i \mid\psi_{\{n\}}\rangle = (-1)^{n_i}\mid\psi_{\{n\}}\rangle. \label{psi_n}
\ee
It follows that 
\bea 
Z_\text{s-f} &=& 
\Tr\Big(\text{e}^{-\beta\mathcal{H}_\text{s-f}}\Big)= 
\sum_{\{n\}}\,\Tr_{\{n\}}
\Big(\text{e}^{-\beta\mathcal{H}_\text{s-f}}\Big)\nonumber\\
&=& \sum_{\{n\}}\,Z_{\text{s-f}\,\{n\}},\label{Z_s-f}
\eea
where $\Tr_{\{n\}}$ stands for the trace within the invariant subspace $\{n\}$.  All invariant subspaces are 
actually degenerate at particle-hole symmetry and in lattices with infinite coordination. Therefore 
$Z_{\text{s-f}\,\{n\}} = Z_*$, $\forall~\{n\}$, so that $Z_\text{s-f} = 2^N\,Z_*$; a degeneracy that cancels out the prefactor in Eq.~\eqn{mapping-2}.

In view of the above result, it is worth briefly discussing the meaning of gauge symmetry breaking.
We observe that, within each subspace, the matrix elements of any gauge variant operator vanish. For instance, 
\ba
\langle \psi'_{\{n\}}\mid \sigma^x_i \mid \psi_{\{n\}}\rangle \!\!&=& \!\!
(-1)^{2n_i} \langle \psi'_{\{n\}}\mid \Omega_i\, \sigma^z_i\,\sigma^x_i \sigma^z_i\,\Omega_i \mid \psi_{\{n\}}\rangle \\
&=& - \langle \psi'_{\{n\}}\mid \sigma^x_i \mid \psi_{\{n\}}\rangle,
\ea
hence $\langle \psi'_{\{n\}}\mid \sigma^x_i \mid \psi_{\{n\}}\rangle=0$. Therefore spontaneous gauge symmetry breaking can be revealed within each invariant subspace only by the long-time behavior of gauge-invariant correlation functions, e.g. $\lim_{t\to\infty}\langle \sigma^x_i(t)\,\sigma^x_i(0)\rangle$. Alternatively, one can add to the Hamiltonian a test field that explicitly breaks the symmetry, and study the response in the limit of vanishing field.    

\subsection{Zero temperature}

We shall start our analysis with a mean-field decomposition. At
zero-temperature, we assume a factorized variational state \be
\ket{\Psi} =
\ket{\text{Ising}}\,\otimes\ket{\text{fermions}}.\label{Psi} \ee 
The Ising wave-function thus corresponds to the ground state of the
Ising model in a transverse field 
\bea \mathcal{H}_\text{Ising} &=& -
\frac{2}{z}\,\sum_{<i,j>}\, J_{ij}\, \sigma^x_i\sigma^x_j
-\fract{U}{4}\,\sum_i\,\sigma^z_i,\;\;\label{H_Ising} \eea
where, assuming translational symmetry,
\bea J_{ij} &=&
\fract{t\sqrt{z}}{2}\,\sum_\sigma \, \bra{\text{fermions}}
c^\dagger_{i\sigma}c^\dagga_{j\sigma}+H.c.
\ket{\text{fermions}}\nonumber\\ & = & J, \qquad \forall\,
<i,j>.\label{J} \eea

In turn, $\ket{\text{fermions}}$ is the ground state of
\bea
\mathcal{H}_\text{fermions} &=& -
\fract{t_*}{\sqrt{z}}\,\sum_{<i,j>}\sum_\sigma\,
\big(c^\dagger_{i\sigma}c^\dagga_{j\sigma}+H.c.\big) .
\label{H_fermions} 
\eea 
Here 
\be 
t_* = t\;\bra{\text{Ising}}\sigma^x_i\sigma^x_j \ket{\text{Ising}}, \label{t_*}
\ee
does not depend on the specific nearest-neighbor bond as we have
assumed translational symmetry. 

It follows that $\ket{\text{fermions}}$ is just the Fermi sea of a
tight-binding Hamiltonian with constant hopping. This implies that
$J$ in Eq.~\eqn{J} has the value $J=8t/3\pi$ independent of
$\ket{\text{Ising}}$. 

The Ising model in Eq.~\eqn{H_Ising} with constant $J$ can be readily
solved on the Bethe lattice, where the mean-field results becomes
exact for $z\to\infty$. In particular one finds that, for any site $i$
\be
\left|\bra{\text{Ising}}\sigma^x_i \ket{\text{Ising}}\right| =
\theta\left(U_c-U\right)\, \sqrt{1-\fract{U^2}{U_c^2}\;}, \label{order-parameter} 
\ee 
where $U_c=8J$. Therefore, for 
$U< U_c $, the mean-field state has a finite order parameter that
vanishes at $U_c$ and above it. In the original model,
Eq.~\eqn{H_s-f}, the finite value of $\langle \sigma^x_i\rangle$
actually implies that the local $Z_2$ gauge symmetry Eq.~\eqn{Z2} is
broken. In fact, we observe that, given the variational wavefunction
$\mid\Psi\rangle$ in Eq.~\eqn{Psi}, the state \be \mid \Psi'\rangle =
\prod_i\,\Big(\sigma^z_i\,\Omega_i\Big)^{m_i}\mid\Psi\rangle, \ee with
$m_i=0,1$, is also solution of the mean-field equations with the same
energy and is orthogonal to $\mid\!\Psi\rangle$. In fact, for $i\not =j$ the average value 
\bea
\langle \Psi\!\mid \Omega_i\,\Omega_j\mid\!\Psi\rangle &=& 
\langle \Psi\!\mid \Omega_i\mid\!\Psi\rangle\langle\Psi\!\mid\Omega_j\mid\!\Psi\rangle 
+ \text{O}\big(z^{-2|i-j|}\big)\nonumber\\
&\sim& \text{O}\big(z^{-2|i-j|}\big),\label{Omega_i Omega_j}
\eea
because the average of any $\Omega_i$ vanishes on the half-filled Fermi sea. Since for any given 
$i$ there are $z^r$ sites $j$ at distance $|i-j|=r$, one readily concludes that $\mid\Psi\rangle$ and 
$\mid\Psi'\rangle$ are indeed orthogonal for $z\to\infty$. Therefore  the mean-field lowest energy state is $2^N$ times degenerate, once more signaling the gauge symmetry breaking.
Such degeneracy exactly compensates the entropic pre-factor in
Eq.~\eqn{mapping-2}, thus providing a physically meaningful
variational estimate of the Hubbard model ground state energy. We note that, on the insulating side where 
$t_*=0$, see Eq.~\eqn{t_*}, the fermionic mean-field Hamiltonian in Eq.~\eqn{H_fermions} vanishes, 
hence any wavefunction $\ket{\text{fermions}}$ solves the variational problem, leading to a degeneracy 
$4^N$ that is only partly compensated by the pre-factor in
Eq.~\eqn{mapping-2}. The net result is a residual entropy $S=N\ln 2$, which physically corresponds to the spin entropy of the hypothetical paramagnetic Mott insulator.

At first sight, the spontaneous symmetry breaking might look wrong by
the Elitzur's theorem.~\cite{Elitzur} In reality, as we mentioned,
this theorem does not apply in infinite-coordination lattices, where
such a symmetry is not impeded from breaking
spontaneously.~\cite{Maslanka} The phase with broken $Z_2$ gauge
symmetry actually corresponds to a metal phase in the Hubbard model
\eqn{Ham-Hub}, while the phase at $U>U_c$ to the Mott insulator.
Thence $U_c = 32/3\pi~D\simeq 3.39~D$ is the critical value of the
Mott transition, which slightly overestimates the actual value
$U_c\simeq 2.97~D$, see Fig.~\ref{phasediagram}. 

We mention that the zero-temperature mean-field solution is equivalent to the
Gutzwiller approximation applied to the Hamiltonian
Eq.~\eqn{Ham-Hub},~\cite{Hvar,Marco-PRB} an approximation that in fact
provides exact variational results in infinite-coordination
lattices.~\cite{Gebhard,mio-dimer} The inclusion of quantum
fluctuations on top of the mean-field solution is equivalent to the
time-dependent Gutzwiller approximation.~\cite{Marco-PRL} Following
Ref.~\onlinecite{Marco-PRB}, if we apply the spin-wave approximation
to the Ising model by writing 
\bea \sigma_i^x &\simeq&
\sin\theta\,\left(2-x_i^2-p_i^2\right)
-\sqrt{2}\;\cos\theta\,x_i,\label{x}\\ \sigma_i^z &\simeq&
\cos\theta\,\left(2-x_i^2-p_i^2\right)
+\sqrt{2}\;\sin\theta\,x_i,\label{z}\\ \sigma_i^y &\simeq&
\sqrt{2}\;p_i,\label{y} \eea where $x_i$ and $p_i$ are conjugate
variables and \be \cos\theta = 
\begin{cases}
U/U_c \;\; & \text{if~} U\leq U_c,\\
1 \;\; & \text{if~} U> U_c,
\end{cases}
\ee
the Hamiltonian \eqn{H_Ising} becomes quadratic 
\be
\mathcal{H}_\text{Ising} \simeq E_0 + \frac{a}{2}\sum_i\, \big(x_i^2+p_i^2\big)
+ \frac{b}{2}\,\frac{2}{z}\,\sum_{<i,j>}\,x_i\,x_j,
\ee
where $E_0$ is the mean-field energy, $a=U_c/2$ and $b=U^2/2U_c$ for
$U\leq U_c$, while $a=U/2$ and $b=U_c/2$ for $U>U_c$.
~\cite{Marco-PRB} The spin-wave spectrum is limited within the energy
window $\omega_\text{m}\leq \omega\leq \omega_\text{M}$, where
$\omega_\text{M} = \sqrt{a(a+b)}$ and 
\be
\omega_\text{m}= \sqrt{a(a-b)}\, .\label{w-min}
\ee 
In other words, the Ising excitations are gapped everywhere except at
the Mott transition where $a=b$, close to which
\be
\omega_\text{m} \simeq \fract{U_c}{2}\,\left| 1 - \fract{U}{U_c}\right|^{1/2}\,.
\ee

\subsection{Finite temperature}
The above mean-field approach can be extended at finite temperature without embarking into delicate issues related to gauge redundancy by simply  exploiting the aforementioned equivalence with the Gutzwiller variational approach.~\cite{Hvar} Within the latter scheme, one assumes for the original Hubbard model Eq.~\eqn{Ham-Hub} the density matrix 
\be
\rho_\text{G} = \mathcal{P}\,\rho_\text{fermions}\,\mathcal{P}^\dagger
= \prod_i\, \mathcal{P}^\dagga_i\,\rho_\text{fermions}\,\mathcal{P}_i^\dagger
,
\label{rhoG}
\ee
where $\rho_\text{fermions}$ is the Boltzmann-Gibbs distribution of free fermions with 
nearest-neighbor 
hopping at half-filling, and the linear operator 
\be
\mathcal{P}_i = \fract{\Phi_{\Uparrow}}{\sqrt{2}}\Big(1+\Omega_i\Big) + 
\fract{\Phi_{\Downarrow}}{\sqrt{2}}\Big(1-\Omega_i\Big),
\ee
with $\Phi_{\Uparrow}$ and $\Phi_\Downarrow$ real variational parameters such that 
$\Phi_\Downarrow^2 + \Phi_{\Uparrow}^2=1$. We 
observe that $1+\Omega_i$ and $1-\Omega_i$ project onto states where site $i$ is singly occupied 
and empty/doubly-occupied, respectively. If we formally transform 
\bea
\mathcal{P}_i \to  \bar{\mathcal{P}}_i &=& 
\fract{\Phi_{\Uparrow}}{\sqrt{2}}\,\bigg(1+(-1)^{n_i}\Omega_i\bigg) 
\mid \Uparrow\rangle \nonumber \\
&& \qquad + \fract{\Phi_{\Downarrow}}{\sqrt{2}}\,\bigg(1-(-1)^{n_i}\Omega_i\bigg)\mid \Downarrow\rangle,\;\label{PG_i}
\eea
where $\sigma^z_i \mid \Uparrow\!(\Downarrow)\rangle = \pm \mid \Uparrow\!(\Downarrow)\rangle$, 
we realize that Eq.~\eqn{PG_i} applied on an electronic wavefunction generates a spin-fermion state belonging to the symmetry invariant subspace with $\{n\} = (n_1,n_2,\dots,n_i,\dots,n_N)$, see 
Eq.~\eqn{psi_n}. 
Replacing $\mathcal{P}_i$ with $\bar{\mathcal{P}}_i$ in Eq.~\eqn{rhoG}
yields a variational density-matrix for the spin-fermion Hamiltonian
Eq.~\eqn{H_s-f} restricted to the invariant subspace $\{n\}$: 
\bea
\rho_G \rightarrow \rho_\text{\{n\}} = \prod_i\,\bar{\mathcal{P}}^\dagga_i \;\rho_\text{fermions}\;
\bar{\mathcal{P}}_i^\dagger. \label{rho_s-f}
\eea

Since at half-filling
$\Tr\Big(\rho_\text{fermions}\,\Omega_i\sigma^z_i \Big) = 0$, it follows that on the Bethe-lattice
with $z\to\infty$, 
\ba
\Tr\Big(\rho_\text{\{n\}}\; \sigma^z_i\Big) &=& \Phi_\Downarrow^2 - \Phi_{\Uparrow}^2\equiv m
\ea
and
\ba
\Tr\Big(\rho_\text{\{n\}}\; \sigma^x_i\, c^\dagger_{i\sigma}\; \sigma^x_j \,c^\dagga_{j\sigma}\Big) &=& 
4\,\Phi_{\Uparrow}^2\,\Phi_\Downarrow^2  \nonumber\\
&&  \Tr\Big(\rho_\text{fermions}\; c^\dagger_{i\sigma}\, c^\dagga_{j\sigma}\Big).
\ea
Both results are independent of $\{n\}$. This entails a renormalized
hopping $t\to 4\,\Phi_{\Uparrow}^2\,\Phi_\Downarrow^2\, t = \big(1 -
m^2\big)\,t$. 

Following the finite-temperature calculations of
Refs.~\onlinecite{Wang-fin-T} and \onlinecite{Matteo-fin-T} obtained
within the Gutzwiller approximation, one has ultimately to minimize
the free-energy density
\bea
F(m) &=& -\fract{4T}{\pi D^2}\,\int_{-D}^D\!\!\!d\epsilon\,
\sqrt{D^2-\epsilon^2}\; \ln\Big(1+\text{e}^{-\beta\,\left(1-m^2\right)\,\epsilon}\Big)\nonumber\\
&& - \fract{U}{4}\,\big(m-1\big) + T\,\ln 2 \label{F(T)}\\
&& + T\,\Big(
\frac{1-m}{2}\,\ln\frac{1-m}{2} + \frac{1+m}{2}\,\ln\frac{1+m}{2}
\Big),\nonumber
\eea
where we recognise that the last term is just the entropy of a spin in
a magnetic field $B = T\tanh^{-1} m$ directed along $z$. The phase
diagram of Eq.~\eqn{F(T)} was calculated by the saddle-point method in
Ref.~\onlinecite{Wang-fin-T} and agrees qualitatively with the exact
one shown in Fig.~\ref{phasediagram}. 

Since the variational free-energy functional, Eq.~\eqn{F(T)}, is the
same for any of the $2^N$ subspaces $\{n\}$, such a degeneracy cancels the entropic pre-factor in 
Eq.~\eqn{mapping-2}, bringing once again a free-energy that is well defined in the thermodynamic limit. The above simple variational calculation is however unable to assess whether the symmetry is preserved or spontaneously broken. 
In Sec.~\ref{dmftfiniteT} we shall instead perform a numerically exact calculation at finite temperature to answer such question.

\section{DMFT solution}

In order to assess the validity of the mean-field approximation, we
now move to the exact solution of the model Eq.~\eqn{H_s-f} by means
of the DMFT using the NRG~\cite{Wilson_RMP,Pruschke_RMP} to solve the
impurity problem defined by
\bea
\mathcal{H}_\text{imp} &=& \sum_{\bk\sigma}\,\epsilon_\bk\,c^\dagger_{\bk\sigma}c^\dagga_{\bk\sigma} \nonumber\\
&& + \sigma^x\, \sum_{\bk\sigma}\,V_{\bk}\big(c^\dagger_{\bk\sigma}d^\dagga_\sigma + H.c.
\big) - \fract{U}{4}\,\sigma^z,\label{H_imp}
\eea
which describes a localized $d$-level that hybridizes with a bath of
conduction electrons. The sign of the hybridization depends on an Ising
degree of freedom $\sigma^x$ whose dynamics is controlled by a field
$U/4$ in the $z$-direction. Within DMFT, the bath parameters
$\epsilon_\bk$ and $V_\bk$ are self-consistently
determined.~\cite{DMFT} Specifically, through the Hamiltonian
Eq.~\eqn{H_imp} we calculate the single-particle Green's function of
the physical electron $\sigma^x\,d^\dagga_\sigma$ in Matsubara
frequencies:
\be
G(i\omega_n) = -\int_0^\beta \!\! d\tau\,\text{e}^{i\omega_n\tau}\; 
\langle T_\tau\Big(\sigma^x(\tau)\,d^\dagga_\sigma(\tau)\;\sigma^x(0)\,d^\dagger_\sigma(0)\Big)\rangle,
\label{G-physical}
\ee
and we require that the self-consistency is reached when 
\be
\sum_\bk\,\fract{V_\bk^2}{i\omega_n - \epsilon_\bk} = t^2\,G(i\omega_n).\label{self-consistency}
\ee

We shall focus hereafter on three dynamical quantities: the Green's function of the physical electron 
defined above, Eq.~\eqn{G-physical}, the one of the auxiliary fermion $d^\dagga_\sigma$,
\be
G_d(i\omega_n) = -\int_0^\beta \!\! d\tau\,\text{e}^{i\omega_n\tau}\; 
\langle T_\tau\Big(d^\dagga_\sigma(\tau)\;d^\dagger_\sigma(0)\Big)\rangle,
\label{G-unphysical}
\ee
and finally the Ising spin Green's function
\be
G_\sigma(i\Omega_\lambda) = -\int_0^\beta \!\! d\tau\,\text{e}^{i\Omega_\lambda\tau}\; 
\langle T_\tau\Big(\sigma^x(\tau)\;\sigma^x(0)\Big)\rangle,
\label{G-sigma}
\ee
where $\omega_n$ and $\Omega_\lambda$ are fermionic and bosonic
Matsubara frequencies, respectively. The corresponding spectral
functions are defined by the analytic continuation to the real axis,
for instance
\[
A(\omega) = -\frac{1}{\pi}\;\Im m\,G(i\omega_n\to \omega+i0^+),
\]
and similarly for $G_d(i\omega_n)$ and $G_\sigma(i\Omega_\lambda)$, to
which we shall associate $A_d(\omega)$ and $A_\sigma(\omega)$,
respectively. 

The NRG calculation have been performed using discretization
parameter $\Lambda=2$, with twist-averaging over $N_z=8$ interleaved
discretization grids. The calculations can be performed taking into
account the spin $SU(2)$ symmetry and the axial charge (isospin)
$SU(2)$ symmetry. The finite-temperature expectation values and
dynamical quantities have been computed by the full-density-matrix
algorithm \cite{peters2006,weichselbaum2007} and the self-energy has
been determined using the approach with an auxiliary Green's function
defined for the physical electron operator. \cite{bulla1998}

In order to reveal the spontaneous symmetry breaking, we shall add to
the Hamiltonian Eq.~\eqn{H_imp} a test field 
\be
\delta\mathcal{H} = -h_x\, \sigma^x, \label{test-field}
\ee
and calculate the average of $\sigma^x$ as $h_x\to 0$.  Moreover, we
shall separately discuss the solution at zero and at finite
temperature.

\subsection{$T=0$ DMFT results}

The impurity model Eq.~\eqn{H_imp} without self-consistency and with a
constant density of states of the bath was studied at zero temperature
in Ref.~\onlinecite{Pierpaolo}. It describes a dissipative two-level
system,~\cite{Leggett-RMP} where the eigenstates of $\sigma^x$ label
the two levels and $\sigma^z$ induces tunneling among them. The
dissipative bath is actually sub-Ohmic so that the tunneling
$\sigma^z$ is irrelevant and the level localizes, i.e. $\langle
\sigma^x\rangle \not = 0$.~\cite{Pierpaolo} This case was argued to
correspond to the metallic phase of the Hubbard model
Eq.~\eqn{Ham-Hub}. The issue is whether this is indeed true, namely if
the localization revealed by $\langle \sigma^x\rangle \not = 0$ 
survives the self-consistency condition Eq.~\eqn{self-consistency}. 

\begin{figure}[t]
\centering
\includegraphics[clip,width=0.48\textwidth]{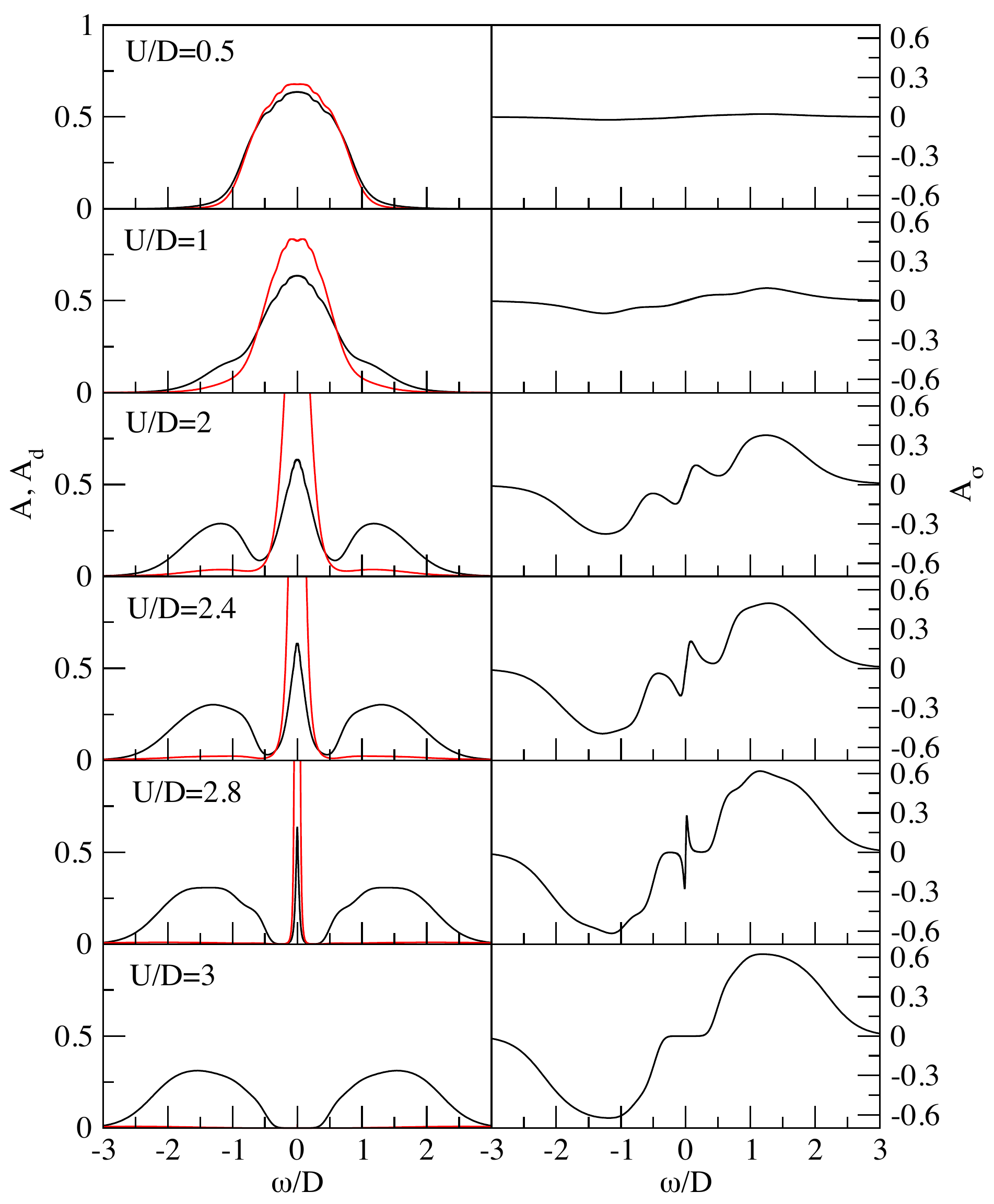}
\caption{(Color online) Left panels: zero-temperature spectral
functions of the physical and auxiliary fermions, $A(\omega)$ and
$A_d(\omega)$ shown as black and red curves, respectively, for a range
of $U$. Right panels: corresponding Ising spin spectral functions
$A_\sigma(\omega)$, which, being the imaginary part of a causal response function, is odd in frequency. 
For $U/D=3$ the system is in the insulating phase.
}
\label{DOS-T=0}
\end{figure}

\begin{figure}[thb]
\centering
\includegraphics[clip,width=0.48\textwidth]{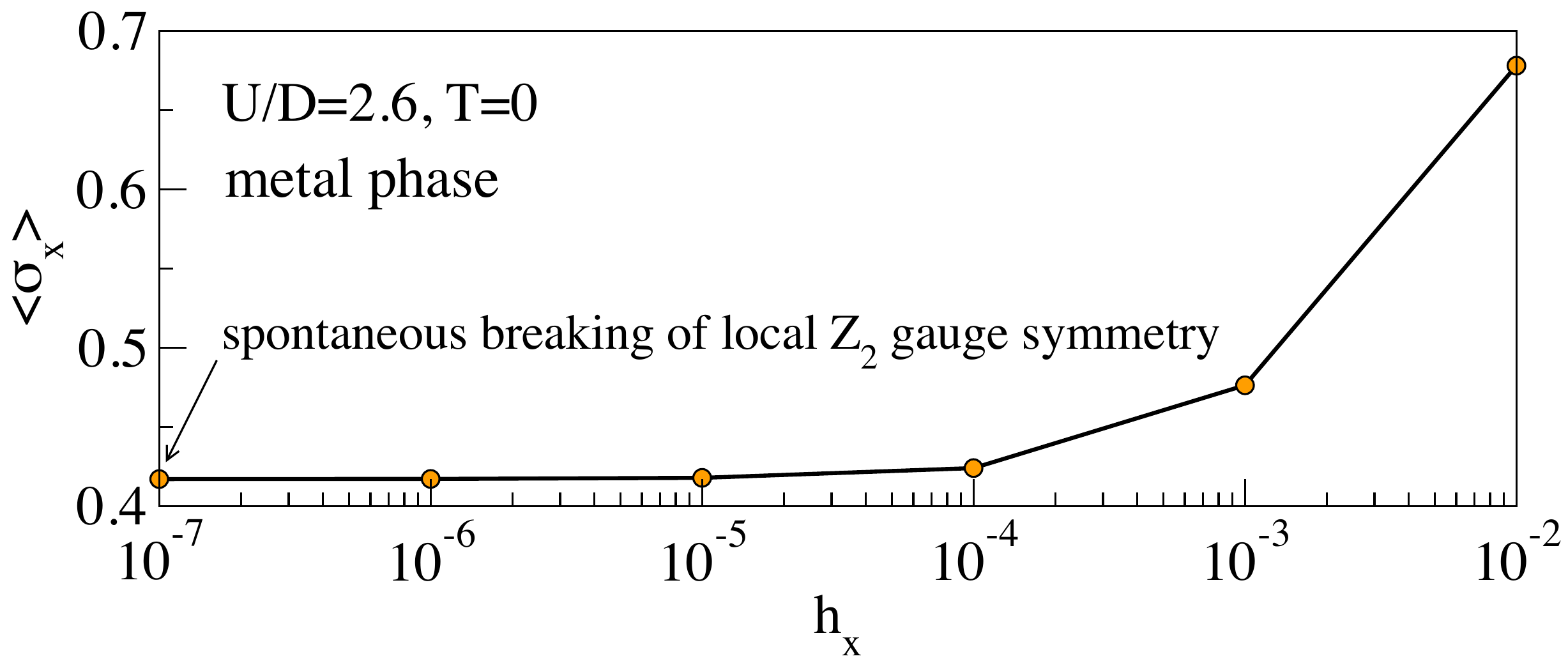}
\caption{(Color online) Ising magnetization $\langle \sigma^x\rangle$
as a function of an external test field $h_x$ in the metallic phase at
$U/D=2.6$ and at zero temperature. We note the finite value as $h_x\to 0$. }
\label{m_x-T=0}
\end{figure}

In Fig.~\ref{DOS-T=0} we show the spectral functions of the physical
and auxiliary fermions, $A(\omega)$ and $A_d(\omega)$, black and red
curves on the left panel, respectively, as well as of the Ising spin
$A_\sigma(\omega)$, right panel, for different values of $U$. Several
features are worth being highlighted. From the physical fermion
spectral function $A(\omega)$ we can locate the Mott transition
between $U/D=2.8$ and $3$, in full agreement with the direct studies
of the Hubbard model.~\cite{DMFT} The manifestations of such a
transition in the behavior of the auxiliary fermion and the slave spin
are quite interesting. The auxiliary fermion has a spectral function
$A_d(\omega)$ resembling that of a resonant model that shrinks
continuously approaching the Mott transition, above which its
weight is concentrated in a $\delta$-peak that is not visible in the
figure. 

The Ising spin shows even more remarkable dynamical features. In the
metallic phase the spectral function $A_\sigma(\omega)$ displays two
peaks at the position of the Hubbard bands but also a linear 
component $A_\sigma(\omega)\sim \omega$ close to zero frequency. In
other words, unlike in the mean-field solution, the Ising spectral
function extends down to zero frequency. The linear low-frequency
behavior occurs in an interval that shrinks as the Mott transition is
approached, and meanwhile the slope increases, just as we would expect
to observe in the imaginary part of the physical-electron charge
response function.  Above the transition, this low energy component
disappears, and a gap opens in the spectrum. This is consistent with
the picture of the Mott insulating state where the charge degrees of
freedom freeze on a high energy scale.

In Fig.~\ref{m_x-T=0} we plot the Ising magnetization $\langle
\sigma^x\rangle$ as a function of an external test field $h_x$, see
Eq.~\eqn{test-field}, in the metallic phase at $U/D=2.6$. We observe
that $\langle \sigma^x\rangle$ approaches a finite value as $h_x\to
0$, thus revealing the spontaneous breaking of the $Z_2$ gauge
symmetry. In the Mott insulating phase $\langle \sigma^x\rangle$
instead vanishes for $h_x\to 0$.  It thus follows that $\langle
\sigma^x\rangle$ is a legitimate order parameter of the
zero-temperature Mott transition in the slave-spin language, finite in
the metal and zero in the insulator, in qualitative agreement with the
mean-field calculation. 

\subsection{$T>0$ DMFT results}
\label{dmftfiniteT}

\begin{figure}[tb]
\centering
\includegraphics[clip,width=0.48\textwidth]{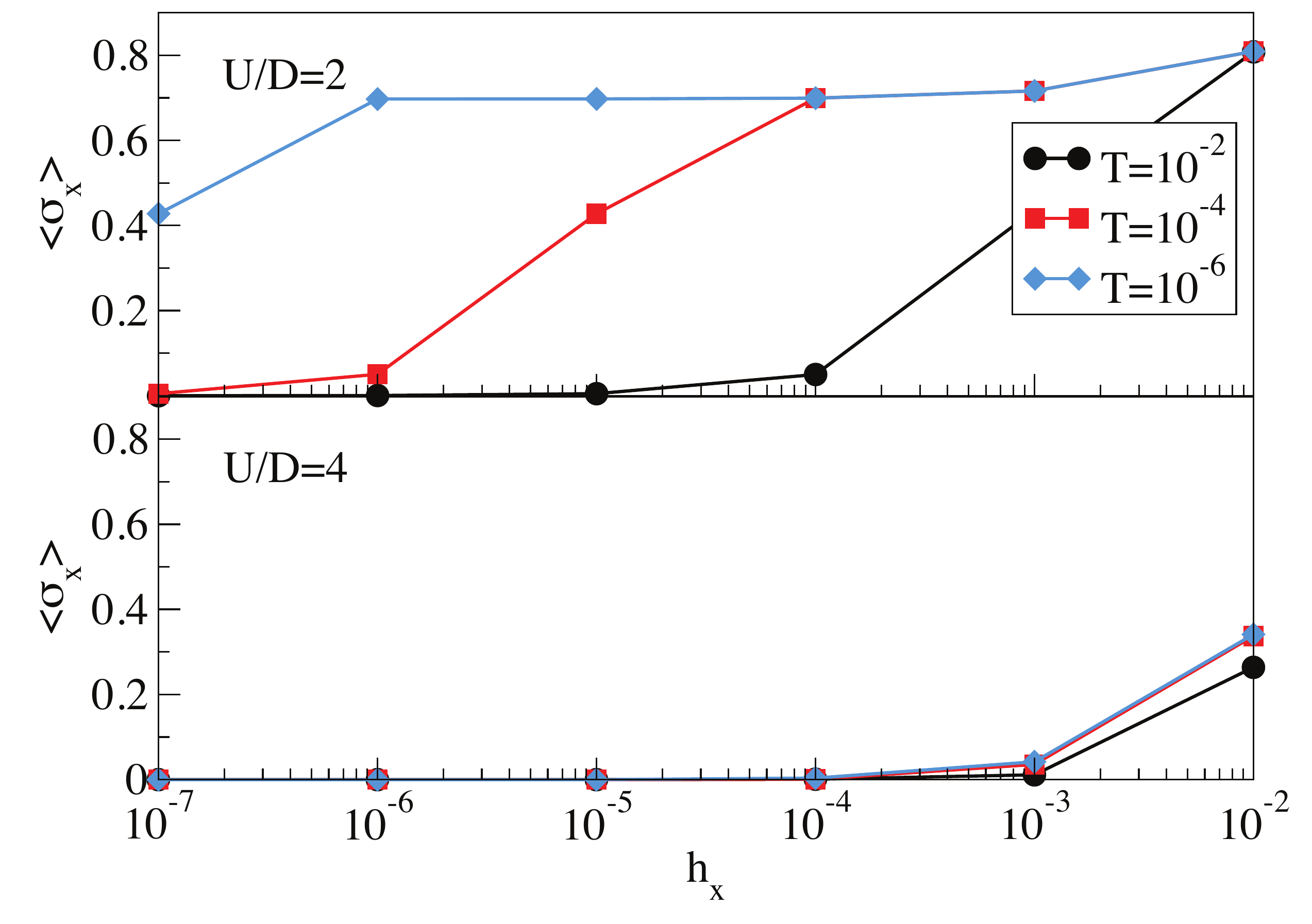}
\caption{(Color online) Ising magnetization $\langle \sigma^x\rangle$
as function of an external test field $h_x$ in the metallic phase at
$U/D=2$ (top panel) and insulating one at $U/D=4$ (bottom panel), for different temperatures. }
\label{m_x-Tnot=0}
\end{figure}

The zero-temperature results demonstrate that the speculations based
on the two-level system Hamiltonian Eq.~\eqn{H_imp} are reliable. We
observe that while at $T=0$ a two-level system coupled to a sub-ohmic
bath localizes, at any $T\not = 0$ it does not, i.e. $\langle
\sigma^x\rangle = 0$, $\forall \,T>0$.~\cite{Leggett-RMP} The recovery
of the gauge symmetry at finite temperature is evident in
Fig.~\ref{m_x-Tnot=0} where we show the slave-spin magnetization $\langle
\sigma^x\rangle$ vs. $h_x$ in the metal (upper panel) for different
temperatures. In the insulator (bottom panel), the magnetization for
$h_x\to 0$ is practically zero at any temperature. In the metal,
however, we observe the gradual onset of a finite order parameter as
$T\to 0$ before $h_x\to 0$. This is further confirmed by the behavior
of the zero-field slave-spin susceptibility, $\chi =
d{\expv{\sigma_x}}/dh_x$. In the metal phase, for fixed and
sufficiently small $h_x$ the susceptibility is increasing as a
power-law $T^{-1}$ in the range $T \gtrsim h_x$ before saturating for
$T \ll h_x$. The saturated value of $\chi(T=0)$ diverges as
$h_x^{-1}$. In the insulator, the susceptibility at fixed $h_x$
is weakly increasing with decreasing $T$ and it saturates on a 
temperature scale of order bandwidth. The saturated value of
$\chi(T=0)$ does not depend on $h_x$ in the insulator.

\begin{figure}[htb]
\centering
\includegraphics[clip,width=0.48\textwidth]{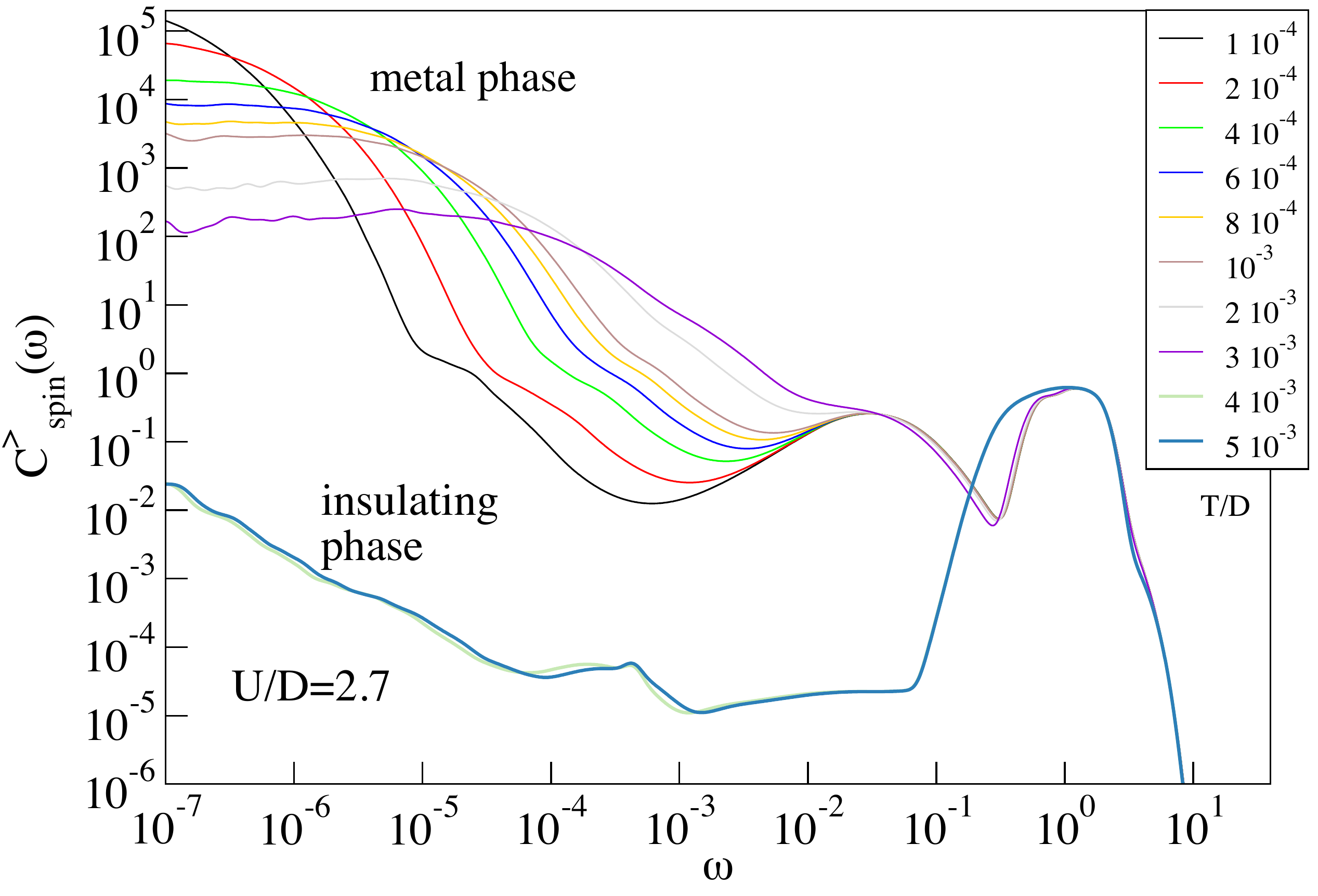}
\caption{(Color online) Retarded Green's function 
$C^{>}_\text{spin}(\omega)$ of the Ising operator $\sigma^x$ at
$U/D=2.7$, i.e. in the metal phase not far from the transition, for
different temperatures. We note the gradual emergence of a
$\delta$-peak at $\omega=0$ on decreasing temperature toward $T=0$.}
\label{Gret}
\end{figure}

The establishment of spontaneous symmetry breaking is also clear in
Fig.~\ref{Gret}, where we plot (at zero test field, $h_x=0$) the
correlation function of the Ising spin, i.e. 
\bea
C^>_\text{spin}(\omega) \!&=&\! 
 \fract{1}{Z}\sum_n\, 
 \text{e}^{-\beta E_n}\! \left| \langle n\mid \sigma^x \mid m\rangle\right|^2 
 \delta\big(\omega - E_m +E_n\big)\nonumber\\
 &=&  \int_{-\infty}^\infty dt\, \text{e}^{i\omega t}\; \langle \,\sigma^x(t)\,\sigma^x(0)\, \rangle,
 \label{G-ret}
 \eea
which is better suited to capture the low-frequency singularity
without the spectral weight cancellation due to broadening that
affects more severely the (odd in frequency) Green's function $A_\sigma(\omega)$. We
observe that $C^{>}_\text{spin}(\omega)$ gradually develops a
$\delta(\omega)$-peak as $T$ decreases, which means that 
\[
\lim_{t\to\infty}\;\lim_{T\to 0}\; \langle \,\sigma^x(t)\,\sigma^x(0)\, \rangle 
\, \rightarrow \, \langle \sigma^x\rangle^2 > 0,
\]
hence the spontaneous symmetry breaking. \footnote{In the NRG, it is
difficult to reliably compute dynamical quantities in the frequency
range $|\omega| \lesssim T$. In spite of this, the tendency towards the
formation of a $\delta$-peak in the $T\to0$ limit can be established.}

\begin{figure}[ht]
\centering
\includegraphics[clip,width=0.48\textwidth]{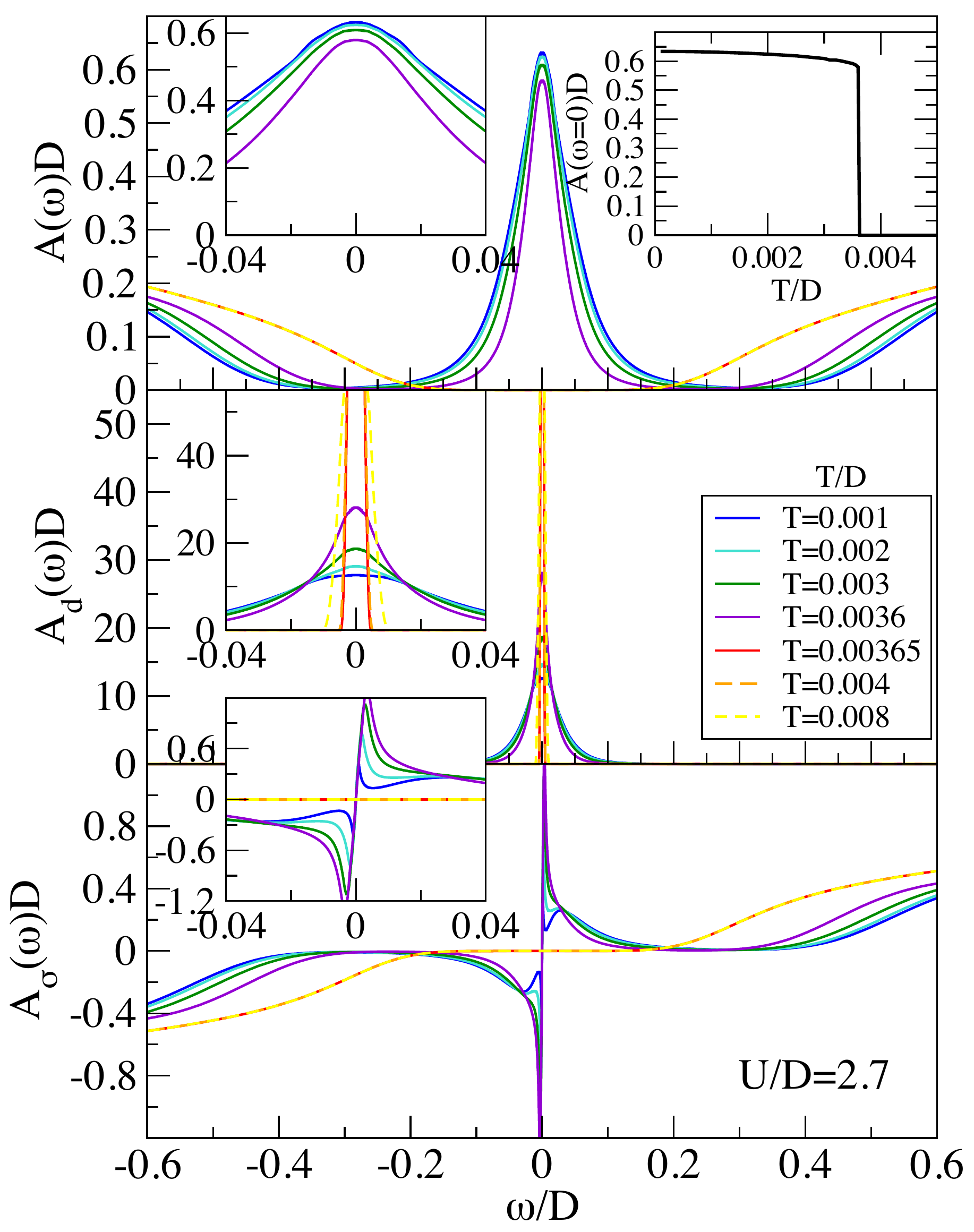}
\caption{(Color online) Spectral functions of the physical electron (top panel), of
the auxiliary fermion (middle panel), and of the Ising spin (bottom
panel) at $U=2.7~D$ upon increasing the temperature across the first
order metal-insulator transition. The insets show the close-ups to the
low-frequency behavior. The right-hand inset in the top panel shows
the density of states at the chemical potential of the physical
electrons at $U=2.7~D$ upon increasing the temperature.}
\label{DOS-Tnot=0}
\end{figure}

In Fig.~\ref{DOS-Tnot=0}(a) we show the temperature evolution of the
spectral functions $A(\omega)$, $A_d(\omega)$ and $A_\sigma(\omega)$
at $U=2.7~D$, when the system is in the metal phase at zero
temperature and, upon heating, crosses the first-order transition and
turns into a Mott insulator, see Fig.~\ref{phasediagram}. The
close-ups to the low-frequency behavior are shown as insets.
We observe that, across the phase transition, the density of states of
the physical electron jumps down from a finite to an extremely small
value, as shown in detail in Fig.~\ref{DOS-Tnot=0} (right inset to top
panel), which is what one also obtains directly from the Hubbard model
Eq.~\eqn{Ham-Hub}.

The novelty here is represented by the behavior of the auxiliary
fermion and slave-spin spectral functions.  We observe that
$A_d(\omega)$ does not change qualitatively across the transition; it
always looks like the spectral function of a resonant level model,
whose width and height simply jump at the transition.

More remarkable is the behavior of the slave-spin spectral function
$A_\sigma(\omega)$. Until the system stays in the metal phase,
$T\lesssim 0.0037~D$, $A_\sigma(\omega)\sim \omega$ at small $\omega$,
with a slope that is practically temperature independent, see
Fig.~\ref{DOS-Tnot=0} bottom panel.  Unlike at the
zero-temperature Mott transition, the interval in which
$A_\sigma(\omega)\sim \omega$ increases with temperature until, above
the first order transition, suddenly disappears; all low energy 
spectral weight is transferred at high-energy, see
Fig.~\ref{DOS-Tnot=0}, and a gap opens.

\begin{figure}[ht]
\centering
\includegraphics[clip,width=0.48\textwidth]{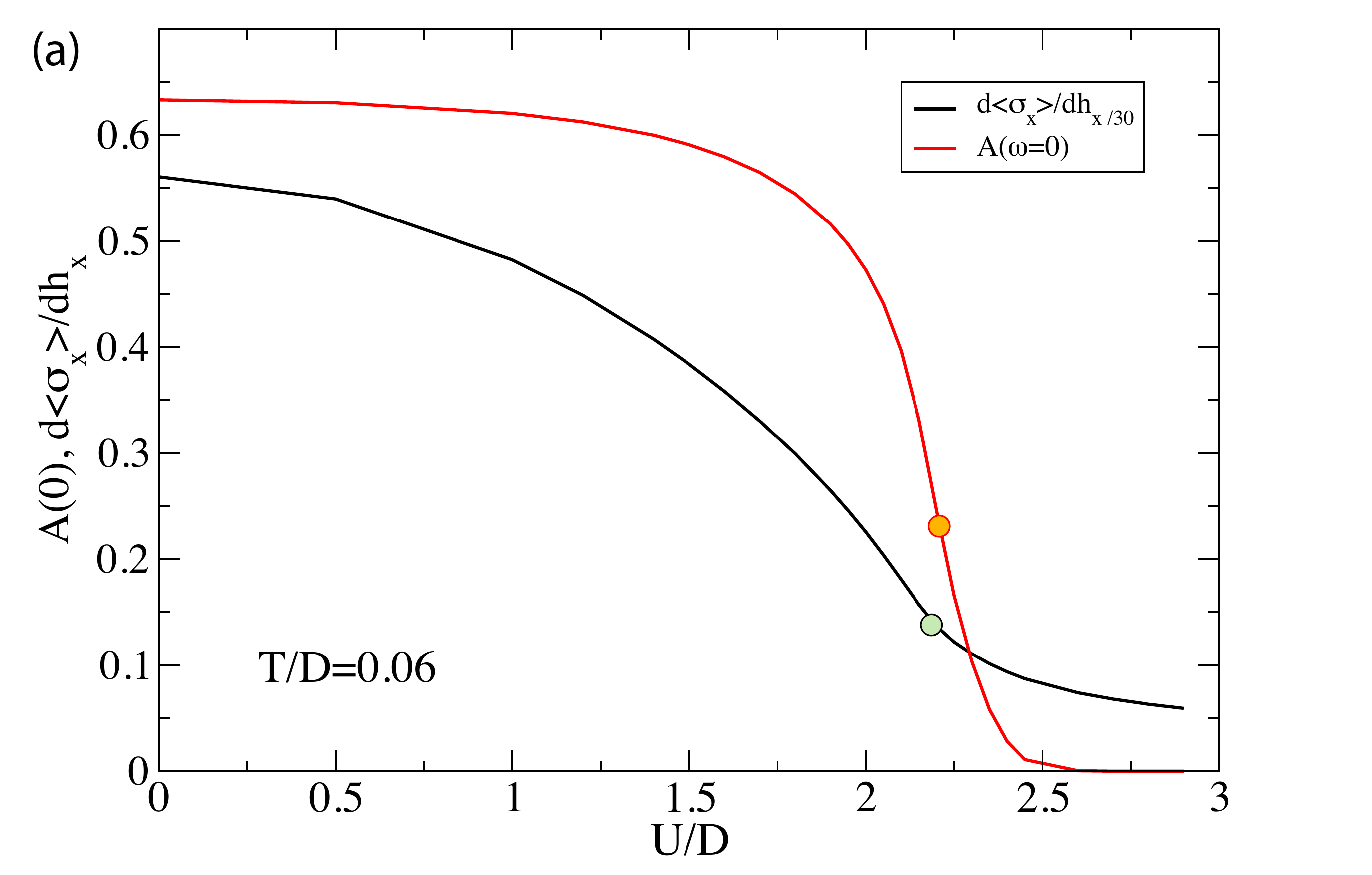}
\includegraphics[clip,width=0.48\textwidth]{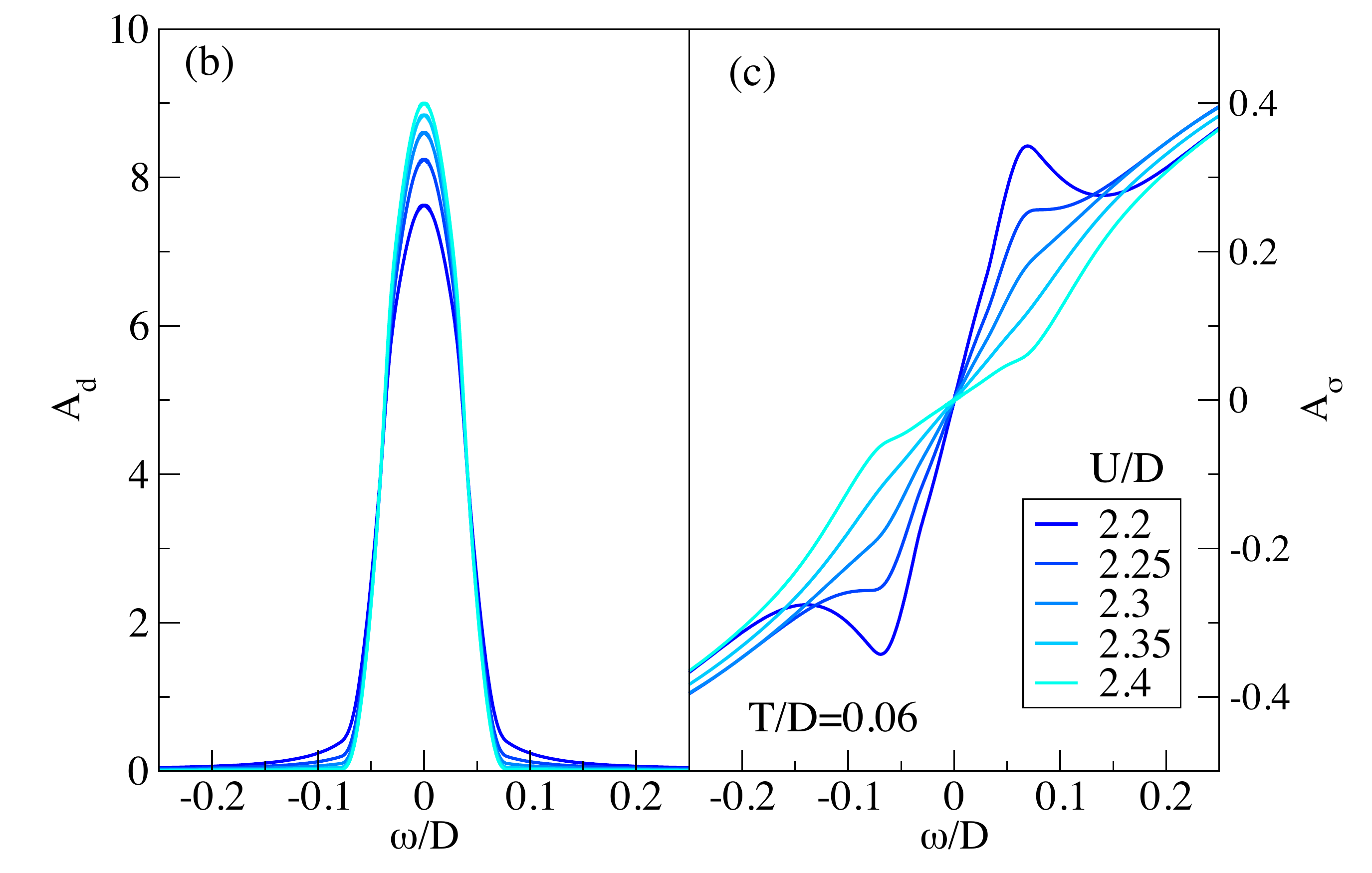}
\caption{(Color online) (a) Physical-electron spectral function intensity at $\omega=0$ and the slave-spin
susceptibility $d\expv{\sigma_x}/dh_x$ at constant temperature $T>T_c$
as functions of $U$. The circles indicate the inflection points on
both curves that may serve as possible definitions of the cross-over
between the bad metal and the Mott insulator.
(b) Auxiliary-fermion and (c) slave-spin spectral functions across the
metal-insulator cross-over.
}
\label{susc1}
\end{figure}

For $T>T_c\simeq 0.026~D$, i.e. above the critical endpoint, see Fig.~\ref{phasediagram}, there is no metal-insulator transition anymore upon increasing $U$; just a smooth
cross-over between a bad metal
(large resistivity, increasing with $T$) and a Mott insulator (large
but finite resistivity, decreasing with $T$). There is no unique
definition of the crossover location, but it is empirically found that
different choices, using either the spectral function, the 
resistivity, or the double occupancy, bring to almost coincident values (e.g., see Fig.~4 in
Ref.~\onlinecite{vucicevic2013}). We have found that in the slave-spin language the
cross-over also reveals itself by 
an inflection point in the
slave-spin susceptibility $d\expv{\sigma_x}/dh_x$, see
Fig.~\ref{susc1}(a).
The evolution of spectral functions across the cross-over is
smooth although we find a notable difference between auxiliary fermions and slave spins. The auxiliary-fermion spectrum changes only qualitatively: with increasing $U$ the width of the peak narrows down  and its height thus raises, see Fig.~\ref{susc1}(b). Since the spin operators of auxiliary and physical fermions map onto each other,  
the above evolution simply represents the gradual formation of localised moments. 
On the contrary, the slave-spin spectrum, which describes charge degrees of freedom, shows 
a more pronounced variation: the linear slope of the 
low-frequency component drops markedly, see
Fig.~\ref{susc1}(c), signalling a rapid cross-over from a gapless to 
a pseudo-gap regime. 

\subsection{Gauge-symmetry breaking's fate at $T>0$}

 The previous results confirm the expectation that, as soon as $T>0$, the Ising order parameter 
 $\langle \sigma^x_i\rangle=0$, in the sense that its expectation value in the presence of a test field 
 $h_x$ vanishes when $h_x\to 0$. To state this in a different manner, we have found that the $2^N$ degeneracy
of each many-body excited eigenstate is not split linearly in
the field strength $h_x$, unlike what happens for the
ground state.
 
We observe that the model Hamiltonian Eq.~\eqn{H_s-f}, which describes 
fermion and spin fields coupled in a gauge-invariant manner, lacks an explicit gauge field. The latter could be made explicit should we decouple the spin-fermion interaction through a Hubbard-Stratonovich transformation involving an auxiliary link-variable that would just play the role of the gauge-field. 
A question we may then ask ourselves is whether, even though the "matter" field, i.e. the Ising field 
$\sigma^x_i$, has vanishing expectation value, still the gauge field has a finite average value, as one would expect in a $Z_2$ gauge-Higgs theory.~\cite{Drouffe-gauge-1,Brezin-gauge,Z2-MC} In order to address this 
question we add to the Hamiltonian Eq.~\eqn{H_s-f} another source field 
\be
\delta  \mathcal{H}_\text{s-f} =  - \frac{h}{\sqrt{z}}\,\sum_{<i,j>}\sum_\sigma\,
\big(c^\dagger_{i\sigma}c^\dagga_{j\sigma}+H.c.\big).\label{source-gauge}
\ee
At finite $h$ the gauge variant operator
$c^\dagger_{i\sigma}c^\dagga_{j\sigma}$ acquires a finite expectation
value, and, accordingly, also $\langle \sigma^x_i \sigma^x_j\rangle$
becomes finite. Within the DMFT mapping onto an impurity model, the
source term Eq.~\eqn{source-gauge} amounts to adding to
Eq.~\eqn{H_imp} the additional term
\be
\delta\mathcal{H}_\text{imp} = 
\sum_{\bk\sigma} \epsilon_{f\bk}\,f^\dagger_{\bk\sigma}f^\dagga_{\bk\sigma} 
+ \sum_{\bk\sigma}\,V_{f\bk}\,\big(f^\dagger_{\bk\sigma} d^\dagga_\sigma + H.c.\big), 
\label{source-imp}
\ee  
which corresponds to hybridizing the impurity with another conduction channel $f$, and impose, 
besides Eq.~\eqn{self-consistency}, also 
\be
\sum_{\bk}\,\fract{V_{f\bk}^2}{i\omega_n - \epsilon_{f\bk}} = h^2\,G_d(i\omega_n).
\label{self-consistency-gauge}
\ee

\begin{figure}[ht]
\centering
\includegraphics[clip,width=0.48\textwidth]{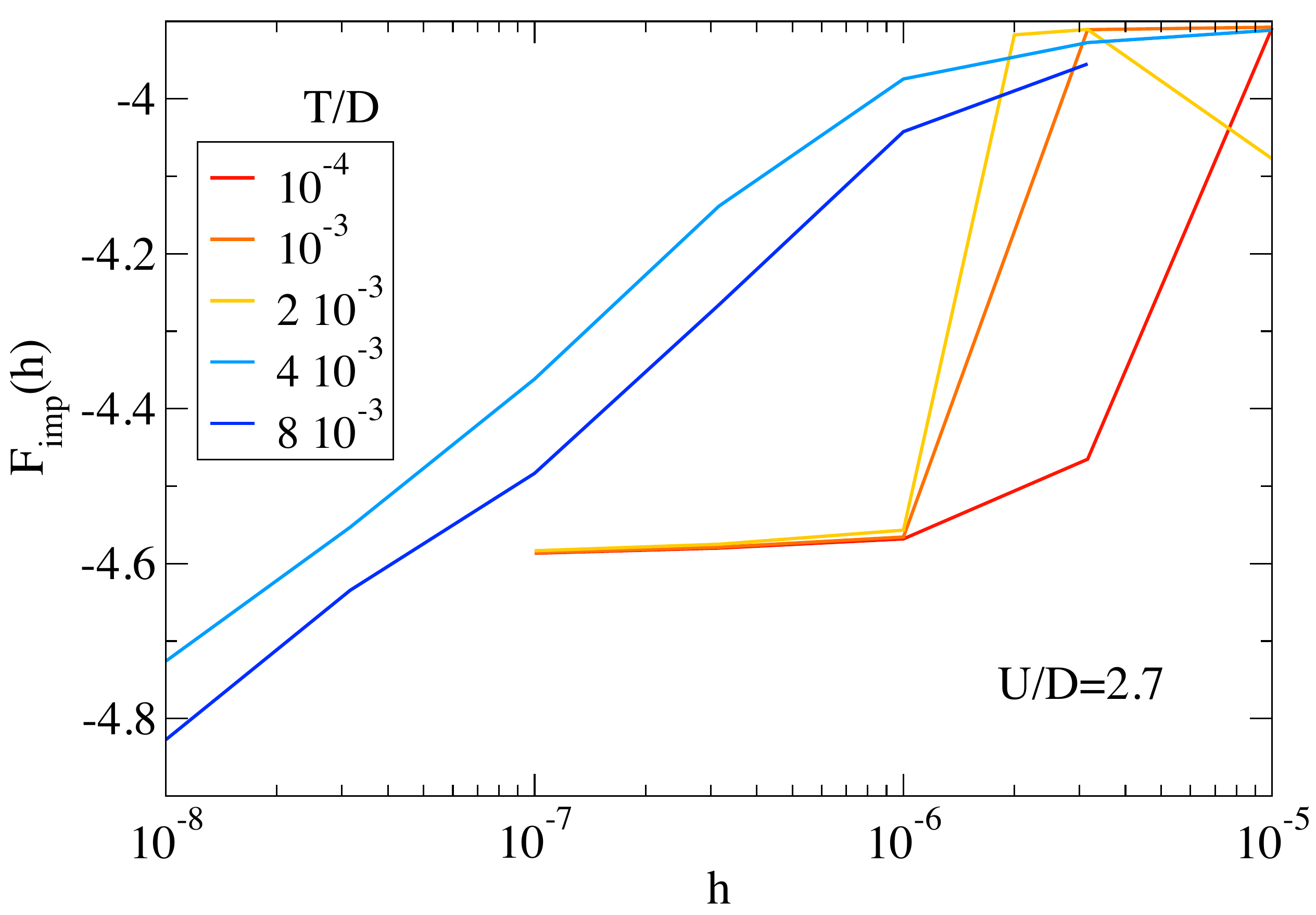}
\caption{(Color online) Impurity free energy for the model with the source term of 
Eq.~\eqn{source-imp}. We
plot the results vs. $h$ for a range of temperatures, both in the
metallic and in the insulating phase.}
\label{src}
\end{figure}

The impurity model $\mathcal{H}_\text{imp} +
\delta\mathcal{H}_\text{imp}$ supplemented by the self-consistency
conditions Eqs.~\eqn{self-consistency} and 
\eqn{self-consistency-gauge} can still be solved by NRG, even though
it now involves two channels of conduction electrons. The solution 
gives access to all local quantities of the original lattice
Hamiltonian $\mathcal{H}_\text{s-f} + \delta\mathcal{H}_\text{s-f}$, 
both static and dynamic. However, in this more complicated situation
we were not able to exploit the cavity method as in
Ref.~\onlinecite{DMFT} and relate local quantities to non-local ones 
so to directly calculate $\langle
c^\dagger_{i\sigma}c^\dagga_{j\sigma}\rangle$. We thence opted for an
indirect route and calculated the free-energy of the impurity model
$F_\text{imp}(h)$ as function of the source field $h$, shown in
Fig.~\ref{src} for different temperatures below (metal) and above
(insulator) the Mott transition at $U=2.7~D$. However, even though the
change that occurs at the finite temperature Mott transition is
observable, the dependence on $h$ is not regular enough to state
unquestionably whether or not the impurity susceptibility diverges in
the limit $h\to 0$. Therefore we cannot give any definite answer to
the question about the finiteness of the gauge-field expectation
value.

\section{Conclusion}

We have shown that the Mott metal-to-insulator transition in infinitely coordinated lattices can be
faithfully reproduced by a model of fermions coupled in a $Z_2$ gauge-invariant manner to Ising spins in a transverse field. We numerically exactly solved such a model by dynamical mean field theory using the numerical renormalization group as impurity solver, and found that gauge-symmetry breaking spontaneously occurs
at zero temperature and corresponds to the metal phase in the original electronic model. 
The zero-temperature metal-to-insulator transition translates in this spin-fermion representation 
to the recovery of the $Z_2$ gauge symmetry, thus endowing the Mott transition of a genuine order parameter. Since we also found that gauge invariance is recovered at any however small finite temperature, we suspect that the Ising magnetization order parameter is closely related to the Fermi surface discontinuity in the quasiparticle distribution, which also disappears at any non-zero temperature. 
While the exact phase diagram agrees qualitatively well with that obtained by mean-field theory, the 
dynamical properties differ substantially. Indeed, contrary to mean-field, we have found that the slave-spin 
spectral function is gapless everywhere in the metallic phase, where it shows a linear dependence 
at low frequency, and becomes gapped only in the Mott insulator. 
This is the case both at zero and at finite temperature, even though in the latter case the transition, 
for $T\lesssim T_c \simeq 0.026~D$, or the crossover, above $T_c$, are not anymore related to the Ising spontaneous magnetization, which is zero for any $T\not = 0$. Our attempts to uncover the behavior of the gauge field, which is hidden in the theory, in order to better characterize the finite temperature phase transition/crossover were unfortunately not conclusive,  even though the dynamical 
behavior of the auxiliary Ising fields notably changes across it.  

\begin{acknowledgments}
R. \v{Z}. acknowledges the support of the Slovenian Research Agency
(ARRS) under Program No. P1-0044. M.~F. thanks G. Martinelli for useful discussions. 
\end{acknowledgments}

\bibliographystyle{apsrev}
\bibliography{./mybiblio}

\end{document}